\documentclass[12pt]{article}

\begin{document}

%%%%%%%%%%%%%%%%%%%%%%%%%%
% Vatche Sahakian's macros

\newcommand{\bb}{\begin{equation}}
\newcommand{\ee}{\end{equation}}
\newcommand{\bbb}{\begin{eqnarray}}
\newcommand{\eee}{\end{eqnarray}}
\newcommand{\diag}{\mbox{diag }}
\newcommand{\Str}{\mbox{STr }}
\newcommand{\Tr}{\mbox{Tr }}
\newcommand{\Det}{\mbox{Det }}
\newcommand{\C}[2]{{\lk [{#1},{#2}\re ]}}
\newcommand{\AC}[2]{{\lk \{{#1},{#2}\re \}}}
\newcommand{\kk}{\hspace{.5em}}
\newcommand{\vc}[1]{\mbox{$\vec{{\bf #1}}$}}
\newcommand{\mc}[1]{\mathcal{#1}}
\newcommand{\del}{\partial}
\newcommand{\lk}{\left}
\newcommand{\ave}[1]{\mbox{$\langle{#1}\rangle$}}
\newcommand{\re}{\right}
\newcommand{\pd}[1]{\frac{\del}{\del #1}}
\newcommand{\pdd}[2]{\frac{\del^2}{\del #1 \del #2}}
\newcommand{\Dd}[1]{\frac{d}{d #1}}
\newcommand{\sech}{\mbox{sech}}
\newcommand{\pref}[1]{(\ref{#1})}

\newcommand
{\sect}[1]{\vspace{20pt}{\LARGE}\noindent
{\bf #1:}}
\newcommand
{\subsect}[1]{\vspace{20pt}\hspace*{10pt}{\Large{$\bullet$}}\mbox{ }
{\bf #1}}
\newcommand
{\subsubsect}[1]{\hspace*{20pt}{\large{$\bullet$}}\mbox{ }
{\bf #1}}

\def\ie{{\it i.e.}}
\def\eg{{\it e.g.}}
\def\cf{{\it c.f.}}
\def\etal{{\it et.al.}}
\def\etc{{\it etc.}}

\def\e{{\mbox{{\bf e}}}}
\def\AA{{\cal A}}
\def\BB{{\cal B}}
\def\CC{{\cal C}}
\def\DD{{\cal D}}
\def\EE{{\cal E}}
\def\FF{{\cal F}}
\def\GG{{\cal G}}
\def\HH{{\cal H}}
\def\II{{\cal I}}
\def\JJ{{\cal J}}
\def\KK{{\cal K}}
\def\LL{{\cal L}}
\def\MM{{\cal M}}
\def\NN{{\cal N}}
\def\OO{{\cal O}}
\def\PP{{\cal P}}
\def\QQ{{\cal Q}}
\def\RR{{\cal R}}
\def\SS{{\cal S}}
\def\TT{{\cal T}}
\def\UU{{\cal U}}
\def\VV{{\cal V}}
\def\WW{{\cal W}}
\def\XX{{\cal X}}
\def\YY{{\cal Y}}
\def\ZZ{{\cal Z}}

\def\sinh{{\rm sinh}}
\def\cosh{{\rm cosh}}
\def\tanh{{\rm tanh}}
\def\sgn{{\rm sgn}}
\def\det{{\rm det}}
\def\trace{{\rm Tr}}
\def\exp{{\rm exp}}
\def\sh{{\rm sh}}
\def\ch{{\rm ch}}

\def\ell{{\it l}}
\def\str{{\it str}}
\def\lp{\ell_{{\rm pl}}}
\def\blp{\overline{\ell}_{{\rm pl}}}
\def\ls{\ell_{{\str}}}
\def\bls{{\bar\ell}_{{\str}}}
\def\bM{{\overline{\rm M}}}
\def\gs{g_\str}
\def\gym{{g_{Y}}}
\def\geff{g_{\rm eff}}
\def\eff{{\rm eff}}
\def\r11{R_{11}}
\def\kel{{2\kappa_{11}^2}}
\def\kten{{2\kappa_{10}^2}}
\def\lpten{{\lp^{(10)}}}
\def\alp{{\alpha '}}
\def\alpe{{{\alpha_e}}}
\def\le{{{l}_e}}
\def\aleff{{\alp_{eff}}}
\def\sqaleff{{\alp_{eff}^2}}
\def\tgs{{\tilde{g}_s}}
\def\talp{{{\tilde{\alpha}}'}}
\def\tlp{{\tilde{\ell}_{{\rm pl}}}}
\def\tr11{{\tilde{R}_{11}}}
\def\wtilde{\widetilde}
\def\what{\widehat}
\def\hlp{{\hat{\ell}_{{\rm pl}}}}
\def\hr11{{\hat{R}_{11}}}
\def\hf{{\textstyle\frac12}}
\def\coeff#1#2{{\textstyle{#1\over#2}}}
\def\CY{Calabi-Yau}
\def\lessapprox{\;{\buildrel{<}\over{\scriptstyle\sim}}\;}
\def\greaterapprox{\;{\buildrel{>}\over{\scriptstyle\sim}}\;}
\def\inbar{\,\vrule height1.5ex width.4pt depth0pt}
\def\IC{\relax\hbox{$\inbar\kern-.3em{\rm C}$}}
\def\IR{\relax{\rm I\kern-.18em R}}
\def\IP{\relax{\rm I\kern-.18em P}}
\def\Z{{\bf Z}}
\def\R{{\bf R}}
\def\One{{1\hskip -3pt {\rm l}}}
\def\sst{\scriptscriptstyle}
\def\osc{{\rm\sst osc}}
\def\lam{\lambda}
\def\lc{{\sst LC}}
\def\pr{{\sst \rm pr}}
\def\cl{{\sst \rm cl}}
\def\D{{\sst D}}
\def\bh{{\sst BH}}
\def\vev#1{\langle#1\rangle}

\input{epsf}

\begin{titlepage}
\rightline{CLNS 02/1799}

\rightline{hep-th/0209179}

\vskip 2cm
\begin{center}
\Large{{\bf Holography with Ramond-Ramond fluxes
}}
\end{center}

\vskip 2cm
\begin{center}
Vatche Sahakian\footnote{\texttt{vvs@mail.lns.cornell.edu}}
\end{center}
\vskip 12pt
\centerline{\sl Laboratory for Elementary Particle Physics}
\centerline{\sl Cornell University}
\centerline{\sl Ithaca, NY 14853, USA}

\vskip 2cm

\begin{abstract}
Starting from the non-linear 
sigma model of the IIB string in the light-cone gauge, we
analyze the role of RR fluxes in Holography. We find that 
the worldsheet theory of
states with only left or right moving modes
does not see the presence of RR fields threading a geometry. 
We use this significant simplification to compute part of the strong
coupling spectrum of the two dimensional NCOS theory. We also reproduce the action of a closed
string in a PP-wave background using this general formalism; 
and we argue for various strategies to find new systems where 
the closed string theory may be exactly solvable.
\end{abstract}

\end{titlepage}
\newpage
\setcounter{page}{1}

\section{Introduction}
\label{intro}

One of the central themes in string theory during recent years is a tantalizing
correspondence between closed and open
string dynamics~\cite{MALDA1}-\cite{ADSLECT}. 
This duality is realized in a myriad of different flavors that yet share certain general 
commonalities. At low energies, the duality often asserts a holographic map between 
gravitational dynamics and certain non-gravitational theories.
The correspondence however necessarily and generally involves the full closed string
theory instead of a low energy truncation to Einstein gravity. Yet technical
problems prevent one from exploring this holographic duality in its full form. One of these problems
has to do with the fact that,
in many settings, the closed string dynamics in question
unfolds in the presence of Ramond-Ramond (RR) fluxes. Understanding
the effects of these fluxes on closed string propagation is hence
of importance to unravelling the duality. 
This issue is of particular interest specially when the dual non-gravitational theory is
a non-commutative open string 
(NCOS) theory~\cite{SWNC}-\cite{SST} \footnote{
Note that the compactified NCOS theory has also a sector of Newtonian gravity~\cite{DGK1}-\cite{DGK2}.}. 
In such scenarios, the holographic duality becomes a map between two
two-dimensional worldsheet theories. We may hope that understanding this map
would teach us fundamental lessons about the
mechanism underlying Holography. 

In this work, we will try to take the first steps in exploring
closed string propagation in RR fluxes. We confine our
discussion to IIB string theory and, after establishing 
a certain general formalism, we will focus on
the case involving the two dimensional NCOS theory and the case of
$AdS_5\times S^5$.

The full light-cone action of IIB string theory in backgrounds of interest was derived in~\cite{SAHRR1}. We 
rewrite here the results in a somewhat more conventional notation. 
We separate the string action into its kinetic piece and two parts labeled by the number of
spinor fields they entail:
\bb\label{action}
\II=\II^{kin}+\II^{(2)}+\II^{(4)}\ .
\ee
The kinetic piece is given by
\bbb\label{Imainkin}
\II^{kin}&\equiv&\frac{1}{2\pi \alpha'} \int d\tau d\sigma\  \lk(2 \sqrt{-h}\ h^{ij} V_i^+ V_j^-
- 4\ \varepsilon^{ij} V_i^+ V_j^- B^{(1)-+}
+\frac{1}{2} \sqrt{-h}\ h^{ij} V_i^a V_{j a} \re.\nonumber \\
&-&\lk.\frac{1}{2} \varepsilon^{ij} V_i^a V_j^b B^{(1)}_{ab}
-2\ i\ V_i^+ \sqrt{-h}\ h^{ij} \bar{\theta} \rho^0 \del_j \theta
+2\ i\ V_i^+ \varepsilon^{ij} \bar{\theta} \rho^1 \del_j \theta\re)\ .
\eee
We have defined 
\bb
V_i^\pm\equiv\del_i x^m e_m^\pm\ ,
\ee
where $e_m^a$ is the vielbein; and
$B^{(1)}_{ab}$ is the NSNS B-field. Spacetime indices are labeled by $m,n,\ldots$,
while tangent space indices are written as $a,b,\ldots$. The light-cone direction is defined as follows
\bb
e^\pm_m=\frac{1}{2} \lk(e^0_m\pm e^d_m\re)\ ,
\ee
with $d$ being an arbitrarily chosen space direction and $0$ denoting the tangent space time index.
Throughout, tangent space indices $a,b,\cdots$
run only over the eight directions transverse to the light-cone; hence, 
we have $a,b,\ldots=1\mbox{ to }9\mbox{ except }d$. The indices $i$ and $j$ will always refer to the
worldsheet coordinates $\tau$ and $\sigma$. Correspondingly, $h^{ij}$ is the worldsheet metric, and
$\varepsilon^{\tau\sigma}=-\varepsilon^{\sigma\tau}=1$. 
Two spinors $\theta_A$, with $A=1,2$, are sixteen component
ten dimensional Majorana-Weyl spinors of the same chirality, and are collected into a doublet 
$\theta\equiv (\theta_1,\theta_2)$. The matrices $\rho^i$ act in this two dimensional space
and may be viewed as defining a worldsheet spinor representation for these fermions.
More details about the spinor representation we are using may be found in Appendix A
and~\cite{SAHRR1}.

The additional pieces of the action involve: `mass terms' for the fermions that we formally write as
\bb\label{I2}
\II^{(2)}\equiv \frac{1}{2\pi \alpha'}\int d\tau d\sigma\  V_i^+ V_j^a \lk( \sqrt{-h}\ h^{ij} \AA_a+\varepsilon^{ij} \BB_a\re)\ ;
\ee
$\AA_a$ and $\BB_a$ are quadratic in $\theta$; and interaction terms quartic in the spinors
given by
\bb\label{I4}
\II^{(4)}\equiv \frac{1}{2\pi \alpha'}\int d\tau d\sigma\  \sqrt{-h}\ h^{ij} V_i^+ V_j^+ \CC\ ,
\ee
with 
\bb\label{cceq}
\CC\equiv \TT^{ab}_{\ \ \ ab} \PP
+\TT^{ca\ b}_{\ \ \ c} \QQ^{(1)}_{ab}
+\SS^{ca\ b}_{\ \ c} \QQ^{(2)}_{ab}
+\TT^{abcd} \RR^{(1)}_{abcd}
+\SS^{abcd} \RR^{(2)}_{abcd}+\JJ\ .
\ee
Two types of quartic terms are involved and are collected into the following combinations\footnote{
Note that our notation for spinors defers from~\cite{SAHRR1}. In particular, $\theta_{old}=\theta_1+i \theta_2$
and $\theta$ is now the doublet $\theta\equiv(\theta_1,\theta_2)$. See Appendix A
for more details.
}
\bb
\TT^{abcd}\equiv (\bar{\theta} \sigma^{ab}\theta) (\bar{\theta} \sigma^{cd}\theta)+
(\bar{\theta} \rho^1 \sigma^{ab}\theta) (\bar{\theta} \rho^1 \sigma^{cd}\theta)\ ;
\ee
\bb
\SS^{abcd}\equiv i (\bar{\theta} \rho^1 \sigma^{ab}\theta) (\bar{\theta} \sigma^{cd}\theta)-
i (\bar{\theta} \sigma^{ab}\theta) (\bar{\theta} \rho^1 \sigma^{cd}\theta)\ .
\ee
These objects have the following symmetries in their index structures
\bb
\TT^{abcd}=\TT^{cdab}=-\TT^{bacd}=-\TT^{abdc}\ ;
\ee
and
\bb
\SS^{abcd}=-\SS^{cdab}=-\SS^{bacd}=-\SS^{abdc}\ .
\ee
In these expressions, $\AA_a$, $\BB_a$, $\PP$, $\QQ_{ab}^{(1)}$,
$\QQ_{ab}^{(2)}$, $\RR_{abcd}^{(1)}$ and $\RR_{abcd}^{(2)}$ 
contain the background supergravity fields in rather elaborate combinations.
These were derived in~\cite{SAHRR1} starting from the superspace formalism, and are reproduced in 
Appendix A in a new more useful notation\footnote{
In passing, let us observe that this action is written by fixing the $SL(2,Z)$ symmetry of
the IIB supergravity theory~\cite{SAHRR1}. As a result, the S-duality group is
broken to a subset given by transformations of the form
\bb
\lk(\begin{array}{cc}
1 & 0 \\
b & 1
\end{array}
\re)\ .
\ee
The remnant symmetry corresponds to shifting the RR axion by $b$.
}. The piece labeled $\JJ$ involves other quartic combinations of the fermions which have not yet been derived and are not relevant to our current analysis\footnote{The explicit form of $\JJ$ will be shortly presented in~\cite{SAHRR1} as well.}.

The action~\pref{action} may be trusted (figuratively speaking) as long as the following conditions are
satisfied~\cite{SAHRR1}: 
(1) All non-bosonic background fields are zero; (2) All background fields are independent 
of the light-cone coordinates; (3) The index structure of the background fields is such that
the light-cone directions `$-$' and `$+$' always appear in pairs, if at all; (4) And the string frame
metric may be cast into a diagonal form. These conditions are satisfied by
most supergravity solutions describing various configurations of $Dp$ branes. Generically,
the light-cone direction is to be chosen such that it is parallel to the worldvolume of one of the D-branes
in a given system. Hence, propagation of closed strings in the vicinity of D-branes may 
readily be studied using
this action. Most of the interaction
terms that appear in~\pref{action} have never been previously explored.
Generically, the implied dynamics is very complex.
Yet, certain aspects, particularly ones relevant to understanding
the holographic duality, may still be unravelled
using various approximation techniques and special settings. 

Before going into any particulars, let us make a few general comments about the structure of our action.
An issue of paramount importance is whether the presence of
quartic terms in the spinors can result in shifting the standard 
fermionic vacuum $\lk<\theta\re>=0$ to a non-zero value. We may expect that in certain
situations, the spinor fields may develop a condensate, much like, for example, in the
Gross-Neveu model~\cite{GROSSNEVEU}. 
Assuming that this is the case, we can easily see that the dynamics of the
bosonic fields $x^m$ describing the embedding of the closed string in the given background
can change dramatically. The fermionic pieces of the action would indirectly  
play the role of sources to the $V_j^a$ field
in the worldsheet theory. Through such a mechanism, RR fields can affect even the leading classical
propagation (\ie\ the ``ground state'') of a closed string 
which is probing a D-brane geometry.

There are two main results in this work that can be summarized as follows. 
For closed string states with only left or right
moving modes on the worldsheet - and in particular for the case of
center of mass motion - we find that
the couplings of the RR fluxes 
to the spinors cancel. This can already 
be seen from the form of our action:
Looking at~\pref{I2} and~\pref{I4}, with the relations
$V^+_\sigma=\pm V^+_\tau$ and $V^a_\sigma=\pm V^a_\tau$, 
and the gauge choice $h^{\tau\tau}=1$ and $h^{\sigma\sigma}=-1$,
it can be observed that both of these pieces vanish. 
The consistency of these conditions and gauge choice with the
worldsheet dynamics is shown in Section 2.
This implies that the complications having to do with a non-trivial
vacuum for the spinors may be circumvented and the dynamics
is determined from the bosonic part of the action\footnote{
Note also that this point is consistent with using perturbations in low
energy supergravity to explore the holographic duality as these
typically correspond to unexcited states of the closed string.
}. We use
this conclusion to compute part of the strong-coupling spectrum of
the two dimensional NCOS theory on a circle. The results agree 
with those of~\cite{VVSlargeM}, but are now presented in the light-cone gauge
with rigorous justification for the needed assumptions.
We then look at the action~\pref{action} for $AdS_5\times S^5$ backgrounds
and we take the PP-wave limit on the worldsheet theory. We argue that
dimensional analysis and the general form of the action immediately
imply that quartic parts in~\pref{I4} are subleading to the kinetic
piece by two powers of the $AdS$ length scale. This general approach can
be used to look for other interesting scaling regimes in different
background geometries.

The outline of the paper is as follows.
In Section 2, we present the classical equations of motion of the closed
string. We distinguish two cases: backgrounds with or without an NSNS
B-field, yet involving RR fluxes. 
In Section 3, we apply our analysis 
to the case corresponding to a holographic duality
between two dimensional NCOS theory on a circle 
and closed strings in a background geometry
involving both NSNS and RR fluxes. 
In Section 4, we present an academic exercise in applying
the technology to the much studied $AdS_5\times S^5$ 
system~\cite{LCADS}-\cite{TADS}; 
and proceed to take the $PP$-wave
limit~\cite{METSPP}-\cite{SRPP} 
to reproduce well-known results that illustrate another mechanism to 
eliminate the non-linearities in the action. In Section 5, we comment on future directions involving
different scenarios, such as integrable worldsheet theories and new and special string backgrounds.

\section{Closed strings and RR fluxes}

In studying the classical dynamics implied by~\pref{action}, we need to subject the system to the constraints
\bb\label{constraint}
T_{ij}\equiv \frac{1}{\sqrt{-h}} \frac{\delta \II}{\delta h^{ij}}=0\ .
\ee
Our action is endowed with two dimensional reparameterization symmetry and scale invariance. 
These allow us to fix the worldsheet metric $h_{ij}$. In general backgrounds, we need
to be careful about this step that is often taken for granted
(see for example~\cite{LCADS,MTADS,TADS}). The equation of motion
for the $V^-$ field is given by
\bb\label{xmneq}
\delta_{V^-} \II=\delta_{V^-} \II^{kin}=0\ .
\ee
For concreteness, let us denote the light-cone directions by $\{ t, y\}$ with corresponding
tangent space labels $\{0, 1\}$. Note that these are isometry directions for our background
by construction. We assume that it is possible to choose coordinates such that
\bb\label{ass}
e^0_t=e^1_y\equiv l\Rightarrow ds^2=l^2\ dt^2-l^2\ dy^2+\cdots\ .
\ee
$l$ may be a function of all the coordinates except $t$ and $y$. 
Assumption~\pref{ass}
is not truly needed, but makes the discussion of the NCOS scenario later
notationally more transparent. We then have
\bb
V_i^\pm=\frac{l}{2} \del_i \lk(t\pm y\re)\equiv l\ \del_i y^\pm\ ,
\ee
which is our definition for $y^\pm$ throughout.
And equation~\pref{xmneq} becomes (using periodic boundary conditions for closed strings)
\bb\label{lccondition}
\del_j\lk(
\sqrt{-h}\ h^{ij} l^2 \del_i y^+
-2\varepsilon^{ij} l^2 \del_i y^+ B^{(1)-+}
\re)=0\ .
\ee
We now need to distinguish two cases: one involving backgrounds without an NSNS
$B$-field; and one that involves a non-zero $B^{(1)-+}$. 

\subsection{Zero NSNS B-field in the light-cone direction}

Let us first assume that we are dealing with a situation where $B^{(1)-+}=0$. 
It is then easy to see that the following conditions
\bb\label{metric1}
\sqrt{-h}\ h^{\tau\tau}=\frac{1}{l^2}\ \ \ ,\ \ \ 
\sqrt{-h}\ h^{\sigma\sigma}=-l^2\ \ \ ,\ \ \ \sqrt{-h}\ h^{\tau\sigma}=0\ ;
\ee
\bb\label{xpl1}
y^+=p^+ \tau\ \ \ \Rightarrow\ \ \  V_\tau^+=l\ p^+\ \ \ ,\ \ \ V_\sigma^+=0\ .
\ee
are consistent with~\pref{lccondition} and fix all the conformal symmetry on the worldsheet. 
Equation~\pref{xpl1}
plays the standard role of eliminating the residual symmetry that is still available
after requiring~\pref{metric1}. Substituting these in~\pref{action}, and rescaling the spinors as in
\bb\label{rescale}
\sqrt{\frac{p^+}{l}}\theta\rightarrow \theta\ ,
\ee
we are lead to the action
\bbb\label{mainp1}
\II&=&\frac{1}{2\pi \alpha'}\int d \tau d\sigma\lk[
-2\ i\ \bar{\theta} \rho^0 \del_\tau{\theta}
+2\ i\ l^2 \bar{\theta} \rho^1 \del_\sigma \theta
+2p^+ \del_\tau y^- \nonumber \re. \\
&+&\lk.\frac{1}{2} \lk( \frac{e_m^a e_{na}}{l^2} \del_\tau x^m \del_\tau x^n
-l^2 e_m^a e_{na} \del_\sigma x^m \del_\sigma x^n\re)
+\del_\tau x^m \AA_a e^a_m+l^2 \del_\sigma x^m \BB_a e^a_m+l^2 \CC
\re]\ ,
\eee
which has properly normalized kinetic terms for the fermions\footnote{Note that
the rescaling~\pref{rescale} does not
introduce additional terms into the action involving derivatives of $l$.}.
The constraints~\pref{constraint} become the two statements
\bbb\label{mainp2}
p^+\lk( \del_\tau \pm \del_\sigma \re)y^- &=&
i\lk(
\bar{\theta} \rho^0 \del_\tau\theta \pm\bar{\theta}\rho^0 \del_\sigma \theta
\re)
-\frac{1}{2} l^2 \CC \nonumber \\
&-&\frac{1}{2} \lk( \del_\tau x^m\pm \del_\sigma x^m\re) \AA_a e^a_m
-\frac{1}{4 l^2} e_m^a e_{na} 
\del_\tau x^m \del_\tau x^n \nonumber \\
&-&\frac{l^2}{4} e_m^a e_{na} \del_\sigma x^m \del_\sigma x^n
\mp \frac{1}{2l^2} e_m^a e_{na} \del_\tau x^m \del_\sigma x^n\ .
\eee
Note that in these two equations, the vielbein appears explicitly, instead of the metric. 
We choose this notation to emphasize that the indices $m$ and $n$ in spacetime 
are restricted to be summed over only the eight directions transverse to the light-cone.

In general, equations~\pref{mainp1} and~\pref{mainp2} describe
a complicated system.
We could choose an ansatz for which the quadratic terms in
the spinors vanish. But with the presence of quartic interactions, 
we should generically expect that
mass terms are generated at the quantum level and a spinor
condensate may develop.
Alternatively, we may focus on BPS-like states of the closed
string, with only left or right moving modes. This would result
in a significant simplification. To make this aspect transparent,
the formalism requires choosing a slightly different and unconventional
gauge on the worldsheet. We will present this in the next subsection, 
where we will be able to consider the case with non-zero $B^{(1)-+}$ in the same
setting. Beyond this,
taming equations~\pref{mainp1} and
~\pref{mainp2} into ones that are computationally manageable 
involves either restricting oneself 
to special backgrounds, either from the outset or
through a judicious scaling regime as in the case of PP-waves geometries; or 
being lucky enough that the corresponding non-linear theory happens
to be integrable. Alternatively, one
can turn around the argument and look for the proper conditions on the background fields
so as to make headway on the problem. We will comment
on these possibilities in Section 4 and the Discussion section. For now, we move onto a more interesting scenario. 

\subsection{Nonzero $B^{(1)-+}$}

Consider next the situation where $B^{(1)-+}$ is
not necessarily zero; \ie\ we have a non-zero NSNS B-field parallel
to the light-cone. 
Equation~\pref{lccondition} may then be solved if 
\bb\label{metric2}
\sqrt{-h}\ h^{\tau\tau}=+1\ \ \ ,\ \ \ 
\sqrt{-h}\ h^{\sigma\sigma}=-1\ \ \ ,\ \ \ 
\sqrt{-h}\ h^{\tau\sigma}=0\ .
\ee
And
\bb\label{xpl2}
\del_{\mp} y^+\equiv \lk(\del_\tau\mp \del_\sigma \re) y^+=0\ \ \ ,\ \ \ e^a_m \del_{\mp} x^m=0\ .
\ee
This means that we allow only for either left moving or right moving modes on the closed string, {\em but 
not both}. It implies that for all background fields, we have
\bb
\del_{\mp} \lk(\mbox{fields}\re)=0\ .
\ee
However, we still have $\del_\mp y^-\neq 0$.
Note that, at this stage, we have fixed the reparametrization symmetry in the standard way by
choosing a worldsheet metric; {\em and} half of the residual conformal symmetry: 
$y^+(\sigma^\pm)$ is still free to be fixed.

All of this formalism goes through as well for the case where
the $B$-field is zero. The difference is that, when $B^{(1)-+}\neq 0$,
this is the only solution we are able to easily write.
It is a mildly entertaining fact
that this reduction in the degrees of freedom makes it `more likely' for a closed string to mimic
open string dynamics. On an open string, right and left moving modes are correlated; on our closed
string, either right or left moving modes are being allowed.
And indeed, it is when the $B$-field is turned on in the time direction that
the dual theory is more than a field theory, but a full (non-commutative) open string theory. 
Yet the solutions given by~\pref{metric2} and~\pref{xpl2} are 
not the only possible
ones. Ours is a condition which is the analogue of the BPS condition for 
a closed string in flat space. We focus on it since it
can describe the center of mass motion of the closed string. 
However, we observe that this discussion suggests that there may be
a dynamical mechanism, summarized by~\pref{lccondition}, 
which results in correlating
the right and left moving modes on the closed string in the presence 
of a B-field for the most general solution to~\pref{lccondition}.

We now proceed to identifying the light-cone momentum by using
\bb
\frac{p^+}{\Sigma}\equiv \frac{\del \II}{\del \del_\tau {{y^-}}}\ .
\ee
We are assuming for convenience
that the $y$ direction is compact of size $\Sigma$; and we are using this
length scale to measure light-cone momentum.
This yields to
\bb\label{ppl}
2 \pi \alpha' \frac{p^+}{2\Sigma}=\lk(l^2\mp B^{(1)}_{ty}\re) \del_\tau {{y^+}}\ ,
\ee
with
$2 B^{-+}=-B_{ty}/l^2$.
Equation~\pref{ppl} is the analogue of $y^+=p^+ \tau$ of the previous subsection. We have now
fixed all of the residual worldsheet symmetries. We summarize these statements again:
\bb
\del_\mp y^+=0\ \ \ ,\ \ \ 
2\pi \alpha' \frac{p^+}{\Sigma}=\lk(l^2\mp B^{(1)}_{ty}\re) \del_\pm y^+\ .
\ee
We then have
\bb
V_\tau^+=2\pi\alpha' \frac{p^+}{2\Sigma} \lk(l\mp\frac{B^{(1)}_{ty}}{l}\re)^{-1}\ ;
\ee
and
\bb
V_\sigma^+=\pm V_\tau^+\ .
\ee
We now substitute all these in~\pref{action}, and, after rescaling the spinors as in
\bb
\sqrt{\frac{p^+}{\Sigma}} \lk(l\mp \frac{B_{ty}}{l}\re)^{-1/2} \theta \rightarrow \theta\ ,
\ee
we obtain the action
\bbb\label{mainpp1}
\II&=&\int d\tau d\sigma\ \frac{p^+}{\Sigma} \del_{\mp} y^-
-i\bar{\theta}\lk(\rho^0\pm\rho^1\re)\del_\pm \theta
 \nonumber \\
&+&\frac{1}{2\pi\alpha'} \lk[
\frac{1}{2} e_m^a e_{na} \del_+ x^m \del_- x^n
-B^{(1)}_{ab} \del_\tau x^m \del_\sigma x^n e^a_m e^b_n
\re]\ .
\eee
The upper/lower choices of
the sign correspond to the possibilities of allowing either
left or right moving modes. Note that only one 
of the two fermions has a kinetic term (see the definitions
of the $\rho$ matrices in Appendix A; the combination $\rho^0\pm\rho^1$ is a projection
operator that picks either spinor $\theta_2$ or spinor $\theta_1$).
The constraints~\pref{constraint} become
\bbb\label{mainpp2}
& &\frac{p^+}{2\Sigma} \frac{l^2}{l^2\mp B^{(1)}_{ty}} \del_\pm y^-
+\frac{1}{8\pi \alpha'}e_m^a e_{na} 
\lk(\del_\tau x^m \del_\tau x^n +\del_\sigma x^m \del_\sigma x^n\re) \nonumber \\
&+&\frac{1}{4} 2\pi \alpha' \CC
+\frac{1}{4} \AA_a e^a_m \del_\pm x^m
-\frac{i}{2}
\bar{\theta} \rho^0 \del_\pm \theta=0\ .
\eee
Note again that $m$ and $n$ are spacetime indices transverse to the light-cone.
We are then left with a single constraint (one for each choice of sign) since
\bb
T_{00}=0\ \ \ \mbox{and}\ \ \ V_\tau^a\mp V_\sigma^a=0\ \ \Rightarrow\ \ 
T_{01}=0\ .
\ee

The interesting fact is that the quartic interaction terms in $\II^{(4)}$ now appear only in the
constraint. The evolution is determined by the action~\pref{mainpp1}, 
which must preserve the
constraint. Hence, if we were to setup initial conditions consistent with the constraint~\pref{mainpp2},
its stability under time evolution is up to~\pref{mainpp1}. This implies that the vacuum 
\bb
\lk<\theta\re>=0\ ,
\ee
taken as a consistent ansatz with the constraint~\pref{mainpp2} for a class of closed
string trajectories, is a stable vacuum. 

We next introduce the light-cone energy $E_{LC}$
\bb\label{elc}
\Pi_{y^+}\equiv \frac{\del\LL}{\del \del_\tau y^+}=
l^2 \lk(1-2 B^{(1)-+}\re) \frac{\del_+ y^-}{2\pi\alpha'}
+l^2 \lk(1+2 B^{(1)-+}\re)\frac{\del_- y^-}{2\pi\alpha'}\equiv \frac{E_{LC}}{\Sigma}+F(\tau,\sigma)\ ,
\ee
where we have used $\lk<\theta\re>=0$.
Center of mass motion would correspond to $F(\tau,\sigma)=0$. And the constraint~\pref{mainpp2}
is the mass-shell condition once one solves~\pref{elc} for $\del_\pm y^-$ and substitutes in~\pref{mainpp2}.
Next, we apply this formalism to the NCOS background, focusing only on center of mass motion of a wound
closed string.

\section{The case of two dimensional NCOS revisited}

\begin{figure}
\epsfysize=3cm \centerline{\leavevmode \epsfbox{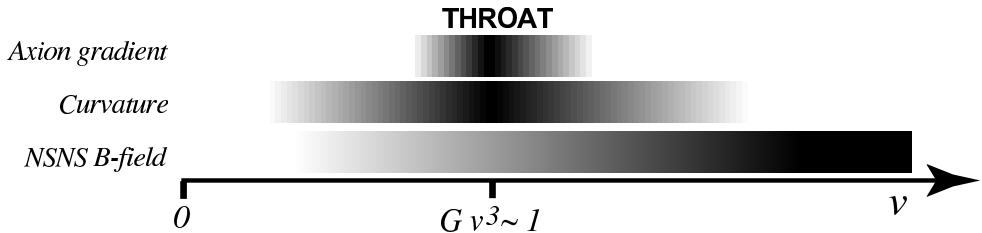}}
\caption{\sl The main features of the NCOS geometry. Darker regions correspond to higher values.
}
\label{fig1}
\end{figure}

The holographic duality of interest to us is the one considered in~\cite{VVSlargeM}. We have
a bound state system of $N$ IIB strings and $M$ D-strings; and we consider
the decoupling scaling limit. The corresponding geometry is given in Appendix B and
involves a non-zero B-field as well as RR gauge fields; one for the IIB axion, and a constant one
from the D-strings. 
We
denote by $t$ and $y$ the directions parallel to the worldvolume of the bound
system, and we use $v$ to denote
the radial direction, which is identified with energy scale in the dual theory.
The geometry involves an interesting throat region at $G v^3\sim 1$. 
The rest of the space is a seven-sphere of size varying with $v$.
The dual theory is two dimensional NCOS theory with string scale $1/(2\pi \alpha_e)=1/(2 \pi \le^2)$,
coupling constant $G$, and D-string charge $M$. The relevant features of the background geometry
are depicted in Figure~\ref{fig1}.

In~\cite{VVSlargeM}, the center of mass motion of a closed string wound in the $y$ direction was studied
after assuming that the RR fields may be ignored. We saw in the previous section that this assumption
is indeed justified. The dynamics explored in~\cite{VVSlargeM} involved, in particular, bounded trajectories
that were used to predict part of the strong coupling spectrum of the NCOS theory. 
Having validated the assumptions of~\cite{VVSlargeM}, we now proceed to reproduce some of the
results, now in the light-cone formalism introduced here. We will find that the
physical conclusions of~\cite{VVSlargeM} are not affected by our choice of a different gauge which happens
to be more useful for understanding the irrelevance of the RR fields to the center of mass
motion. In this regards, we will not present all the details of
the NCOS theory and refer instead the
reader to~\cite{VVSlargeM} for a more complete exposition to the subject.

We consider a closed string wound in the light-cone
direction $y$
radially infalling towards $v=0$. 
Given that one is allowed left or right moving modes only, 
we need to choose the light-cone momentum in such a way as to correlate it with 
the desired winding number if we do not desire to consider the case with additional
momentum along the closed string. Applying $\del_\mp$ to the constraint equation~\pref{mainpp2},
we find that $y^-=f_1(\sigma^+)+f_2(\sigma^-)$. Applying $\del_\mp$ to~\pref{elc} with
$F(\tau,\sigma)=0$ (\ie\ the case of center of mass dynamics), 
we find that $\del_\mp^2 y^-=0$. Hence, we write
\bb\label{Kdef}
y^-=-\frac{2 K}{\Sigma} \sigma^\mp +g(\sigma^\pm)\ .
\ee
The two choices of signs we have for left or right moving modes translate now to two possible 
orientations for the winding: parallel or anti-parallel to the B-field.
We will map the constant $K$ onto winding number soon. And $g(\sigma^\pm)$
is fixed by~\pref{elc}.

For center of mass motion, the classical dynamics is entirely determined by the constraint~\cite{VVSlargeM}.
This now becomes the statement
\bb\label{gvv}
\frac{1}{8} g_{vv} \lk(\del_\pm v\re)^2+\frac{p^+}{2\Sigma} 
\lk(
(2\pi\alpha')^2 \frac{E_{LC}}{\Sigma} \frac{l^2}{l^4-{B^{(1)}_{ty}}^2}+
(2\pi\alpha') \frac{l^2}{l^2\pm B^{(1)}_{ty}} \frac{2 K}{\Sigma}
\re)=0\ ,
\ee
where we have used the fermion vacuum $\lk<\theta \re>=0$.
Denoting the winding number of the closed string along the $y$ direction by $\omega>0$,
while keeping track of orientation through the two choices of signs, we must then have
\bb\label{ko}
K=-\frac{\omega}{2} \Sigma\ ,
\ee
The factor of 2 arises from the definition of the light-cone directions $y^\pm$,~\pref{Kdef},
and by demanding $y\sim y+\Sigma$. 
We have two more constants in the problem, $p^+$ and $E_{LC}$. 
We identify $E_{LC}$ with the energy scale $Q_E$ measured in
the NCOS theory
\bb\label{elcqe}
E_{LC}=Q_E=-\frac{E}{4\pi\alpha_e} \Sigma\ ,
\ee
with $E$ corresponding to a dimensionless measure of energy conveniently
introduced in~\cite{VVSlargeM}. Given the dispersion relation $2 E_{LC} p^+=Q_E^2$, 
with $E_{LC}=Q_E+p^y$ and $p^+=(Q_E-p^y)/2$, this fixes
$p^+=Q_E/2$, which corresponds to $p^y=0$ in the `unboosted' frame.
Indeed,
this is the scenario that was studied in~\cite{VVSlargeM},
where there was no momentum $p^y$ along the winding string.  

The dynamics dictated by~\pref{gvv}
is then equivalent to a particle in one dimension moving in the potential
(see Appendix B for details on the background geometry in question)
\bb\label{pot}
V=\frac{\omega}{2\ G\ \Sigma^2} v
\frac{E\sqrt{1+Gv^3}+\sqrt{G v^3} \lk(2\omega\pm E\re)}{\sqrt{1+G v^3}\pm\sqrt{G v^3}}=
-\frac{1}{2} \lk(\del_\pm v\re)^2\ .
\ee
The motion of the closed string has the same qualitative features as that of~\cite{VVSlargeM}.
In particular, there are identical bounded and scattering
processes; and for bounded dynamics, the turning points are
$v=0$ and  $v=v_c$ given by
\bb
G v_c^3=\frac{E^2}{4 (\omega^2\pm \omega E)}\ .
\ee
as in~\cite{VVSlargeM}. 
This agreement is a check on the
consistency of the formalism and normalizations; 
we have one arbitrary scale but two conditions to satisfy and,
after determining~\pref{ko}, a solution is possible.

Applying the Bohr-Sommerfeld quantization for the bounded dynamics scenario
\bb
\II= \int d\tau d\sigma\ \frac{p^+}{\Sigma} \del_\mp y^-=\pi N\ ,
\ee
we obtain the statement
\bb
\frac{p^+ \omega}{\pi} \Delta \tau=N\ \ \ \mbox{for   }N\gg 1\ ,
\ee
where $\Delta\tau$ is the proper time for a bounce, and $N$ is the level number.
Finding the spectrum then involves integrating~\pref{pot} for $\Delta \tau$ between
the turning points $v=0$ and $v=v_c$, 
which is now slightly more delicate. The following chain of change of variables are very useful 
\bb
G v^3=\sinh^2 t\ ;
\ee
followed by
\bb
e^t=\sqrt{y+1}\ .
\ee
The integral's bounds are
\bb
y_1=0\ \ \ ,\ \ \ 
y_2=\lk \{
\begin{array}{c}
\frac{-E}{\omega+E}>0\ \ \ \mbox{for positive winding} \\
\frac{-E}{\omega}>0\ \ \ \mbox{for negative winding}
\end{array}
\re.
\ee
We then find the proper times for positive and negative winding orientations given by
\bb
\Delta\tau^+=\frac{a_0 \le}{|E|} 
\lk(\frac{|E|G}{E+\omega}\re)^{1/3}
\lk(
\mbox{ }_2F_{1}\lk(-\frac{5}{6},\frac{1}{3},\frac{5}{6},\frac{E}{E+\omega}\re)
+\mbox{ }_2F_{1}\lk(\frac{1}{6},\frac{1}{3},\frac{5}{6},\frac{E}{E+\omega}\re)
\re)\ ;
\ee
\bb
\Delta\tau^-=\frac{a_0 \le}{|E|} 
\lk(\frac{|E|G}{\omega}\re)^{1/3}
\lk(
\mbox{ }_2F_{1}\lk(-\frac{1}{3},\frac{1}{3},\frac{5}{6},\frac{E}{\omega}\re)
+\mbox{ }_2F_{1}\lk(\frac{1}{3},\frac{2}{3},\frac{5}{6},\frac{E}{\omega}\re)
\re)\ ,
\ee
where $\mbox{ }_2F_1$ is a well-known Hypergeometric function,
and
\bb
a_0\equiv \frac{2^{11/3} \pi^{3/2} \Gamma(4/3)}{\Gamma(5/6)}\ .
\ee
The spectrum then becomes
\bb
\le Q_E\simeq \frac{N^3}{G \omega^2} \frac{\alpha_e}{\Sigma^2}\ \ \ \ 
\mbox{for positive winding}\ ;
\ee
and
\bb
\le Q_E\simeq \frac{N^{3/2}}{\sqrt{G \omega}} \sqrt{\frac{\le}{\Sigma}} \ \ \ 
\mbox{for negative winding}\ ;
\ee
with $N \gg 1$.
These are indeed identical to the results of~\cite{VVSlargeM}, even though the steps which 
we used to arrive at it are slightly different. 
We conclude that, despite the presence of RR fluxes, we are able to 
compute part of the strong coupling spectrum of the two dimensional NCOS theory to leading
order in inverse D-string charge $1/M$.

It is instructive to briefly look at the mass terms~\pref{I2} for this NCOS geometry.
The geometry depicted in Figure~\ref{fig1} exhibits an interesting throat region encoded in the metric
and, most interestingly, in the axion field. Indeed, when one substitutes the geometry in question
into~\pref{Aa} and~\pref{Ba}
and truncates to the infalling string ansatz, 
one finds that only two terms are non-zero: one proportional to axion
flux $\del\chi$ in~\pref{Aa}, and one proportional to axion-NSNS-B-field term $N^{-+}_{\ \ \ \ a}$ in~\pref{Ba}.
Both terms are proportional to the same spinor combination $\bar{\theta}\rho^0 \rho^1 \theta$.
It is then most likely that the effect of the throat would be felt on the worldsheet through the worldsheet
spinors becoming massive or massless as the string passes through the region $Gv^3\sim 1$. 
Correspondingly, the excited spectrum of the closed string would very much be sensitive to the throat.

\section{Strings on AdS and PP-wave backgrounds}

We next look at another mechanism by which the quartic interaction terms may disappear
from the dynamics. The starting point is to consider a closed string moving in the
vicinity of a stack of D3 branes.
The $AdS_5\times S^5$ D3 brane geometry in the coordinates used by~\cite{BMN} appears as
\bb\label{AdSmetric}
ds^2=R^2\lk(
\cosh^2\rho\ dt^2
-d\rho^2
-\sinh^2\rho\ d\Omega_3^2
-\cos^2\theta\ d\psi^2
-d\theta^2
-\sin^2 \theta\ d{\Omega'}_3^2
\re)\ ;
\ee
with a five-form RR flux given by
\bb\label{AdSflux}
G_{t\rho\theta_1\phi_1\phi_2}= 4 R^4 \cosh\rho\ \sinh^3 \rho \sqrt{|g_{\Omega_3}|}\ \ \ ,\ \ \
G_{\theta\psi\theta_2\phi_3\phi_4}= 4 R^4 \cos\theta\ \sin^3 \theta \sqrt{|g_{{\Omega'}_3}|}\ .
\ee
The coordinates $\theta_1$, $\phi_1$, and $\phi_2$ parameterize the three-sphere
$\Omega_3$ of volume $\sqrt{|g_{\Omega_3}|}$; 
and $\theta_2$, $\phi_3$, and $\phi_4$ parameterize the three-sphere
${\Omega'}_3$ of volume $\sqrt{|g_{{\Omega'}_3}|}$. 
We separate the index structure for the two spaces we are dealing with:
$a,b,\cdots$ shall refer to tangent space indices in $AdS_5$; while indices
$d,e,\cdots$ shall refer to tangent space indices on the $S^5$. We hope
that there will be no confusion between the $\theta_1$ and $\theta_2$ used as spacetime
coordinates and the spinors in the worldsheet theory labeled perversely with the same
letters.
We then write
\bb
R_{ab}=\frac{4}{R^2} \eta_{ab}\ \ \ ,\ \ \
R_{de}=-\frac{4}{R^2} \eta_{de} \ .
\ee
We summarize our notation:
\bb
\lk\{e^a_m\re\}\sim \lk\{
e^0_t,e^1_\rho,e^2_{\theta_1},e^3_{\phi_1},e^4_{\phi_2}
\re\} \in AdS_5
\ee
and
\bb
\lk\{e^d_m\re\}\sim \lk\{
e^5_\theta,e^6_\psi,e^7_{\theta_2},e^8_{\phi_3},e^9_{\phi_4}
\re\}\in S^5
\ee
We are then to substitute~\pref{AdSmetric} and ~\pref{AdSflux} in~\pref{action}.  
We choose the light-cone direction to be parallel to the D3 brane worldvolume. 
In our parameterization, we pick the $0-2$ (\ie\ the $t$ and $\theta_1$) directions in tangent space
\bb
A^\pm=\frac{1}{2} \lk( A^0\pm A^2\re)\ ,
\ee
where $A^a$ is some arbitrary background field.
After some unpleasant work, we are lead to the following kinetic term
(ignoring the bosonic part)
\bb\label{AdSIkin}
\II^{kin}_{AdS}=-\frac{i R}{2\pi\alpha'} \int d\tau d\sigma \lk(
\frac{\del_i t}{e^t_0}+\frac{\del_i \theta_1}{e^{\theta_1}_2}\re)
\lk(
\sqrt{-h} h^{ij} \bar{\theta} \rho^0 \del_j \theta -\varepsilon^{ij} \bar{\theta} \rho^1 \del_j \theta
\re)\ .
\ee
The mass terms for the spinors are given by
\bbb\label{AdSI2}
\II^{(2)}_{AdS}&=&\frac{i R}{2\pi\alpha'}\int d\tau d\sigma
\lk(
\frac{\del_i t}{e^t_0}+\frac{\del_i \theta_1}{e^{\theta_1}_2}\re)
\lk(
2 \sqrt{-h} h^{ij} \frac{\del_j x^m}{e^m_d} 
\lk(\bar{\theta} \rho^0 \rho^1 \sigma^{134d} \theta\re) \re. \nonumber \\
&-&\lk. \frac{1}{48} \sqrt{-h} h^{ij} \frac{\del_j x^m}{e^m_h}
\lk(\bar{\theta} \rho^0 \rho^1 \sigma^{defg} \theta\re) \varepsilon^{S^5}_{hdefg}
+\frac{\varepsilon^{ij}}{2} \frac{\del_j x^m}{e^m_c}
\lk(\bar{\theta} \sigma^{ab} \theta\re) \varepsilon^{AdS}_{cab} \re. \nonumber \\
&+&\lk. \frac{\del_j x^p}{e^p_k} \omega_{k,ab}
\lk(
\sqrt{-h} h^{ij} \lk(\bar{\theta} \rho^0 \sigma^{ab} \theta\re)
-\varepsilon^{ij} \lk(\bar{\theta}  \rho^1 \sigma^{ab} \theta\re)
\re)
\re)\ .
\eee
$\omega_{k,ab}$ is the connection of the string frame metric, with the index $k$
summed over {\em all} tangent space indices. 
$\varepsilon^{S^5}_{5defg}$ is the antisymmetric form on $S^5$, and
$\varepsilon^{AdS}_{1ab}$ is the antisymmetric form on the $\rho-\phi_1-\phi_2$ patch.
And there are quartic terms in the spinors
\bb\label{AdSI4}
\II^{(4)}_{AdS}=\frac{1}{2\pi\alpha'}\int d\tau d\sigma {\sqrt{-h} h^{ij}}
\lk(
\frac{\del_i t}{e^t_0}+\frac{\del_i \theta_1}{e^{\theta_1}_2}\re)
\lk(
\frac{\del_j t}{e^t_0}+\frac{\del_j \theta_1}{e^{\theta_1}_2}\re)
\lk(
\frac{9}{256} \TT^{de}_{\ \ \ de}-\frac{101}{768} \TT^{ab}_{\ \ \ ab}
\re)\ .
\ee
We expect no contribution from the unknown quartic terms denoted by $\JJ$ in equation~\pref{cceq}.
This is because these terms involve spinors with four or two units of U(1) charge; and the action must be
neutral. On the other hand, the only spacetime fields which carry a balancing charge (the field strengths
for the RR complex scalar and the RR 2-forms) are zero in the AdS background. Hence, from symmetry,
we expect that $\JJ=0$ in equation~\pref{cceq} for the AdS geometry. 
The reader is referred to~\cite{SAHRR1} for more details.
Note also that we have not fixed the worldsheet symmetries to allow for comparison with different
conventions. Partly because of this, 
the action appears elaborate. Another reason is that we used global $AdS$ coordinates. It is
more esthetically pleasing to write the action in Poincar\'{e} coordinates~\cite{LCADS,MTADS,TADS}. But our motivation is to
use it to take the PP-wave limit, and this form is more suitable for this purpose.

Following~\cite{BMN}, we introduce the coordinates
\bbb\label{coord1}
t=x^{\widetilde{+}} +\frac{x^{\widetilde{-}}}{R^2}\ \ \ &;&\ \ \ 
\rho=\frac{r}{R} \ \ ;\nonumber \\
\theta=\frac{y}{R}\ \ \ &;&\ \ \ 
\psi=x^{\widetilde{+}}-\frac{x^{\widetilde{-}}}{R^2}\ .
\eee
Note that the directions $\widetilde{+}$ and $\widetilde{-}$ are different
from the light-cone direction $+$ and $-$. 
This situation arises since our action can be used only if the light-cone chosen in the gamma 
matrix algebra
is parallel to the D3-brane. Whereas the one of interest in the PP-wave limit picks a spacelike cycle on
the $S^5$, a direction transverse to the D3 branes. We then regard~\pref{coord1}
as simply a convenient coordinate change in the bosonic fields to facilitate the process
of taking the PP-wave limit $R\rightarrow \infty$~\cite{BMN}. 

We substitute the new coordinates in~\pref{AdSIkin}-\pref{AdSI4}, 
and we take the $R\rightarrow\infty$ limit; we
arrive at the action
\bbb
\II_{PP}&=&\frac{i R}{2\pi\alpha'}\int d\tau d\sigma \del_i x^{\widetilde{+}} \lk[
\frac{1}{2}\sqrt{-h}\ h^{ij} \lk(
\bar{\theta}\rho^0 \rho^1 \sigma^{5789}\theta+4 \bar{\theta} \rho^0 \rho^1 \sigma^{1634} \theta
\re) \del_j x^{\widetilde{+}} \nonumber \re.\\
&-&\lk. \sqrt{-h} h^{ij} \bar{\theta} \rho^0 \del_j \theta
+\varepsilon^{ij} \bar{\theta} \rho^1 \del_j \theta 
\re]+\II^{bos}_{PP}\ ,
\eee
where $\II^{bos}_{PP}$ is the bosonic part, which is rather trivial.
Using the self-duality of the gamma matrix $\sigma^{abcde}$, we find 
\bb
\ast{{\sigma}}^{abcde}=+\sigma^{abcde}\Rightarrow
\sigma^- \sigma^{1634}=\sigma^- \sigma^{5789}\ .
\ee
We then finally get to 
\bb
\II_{PP}=i R p^{\widetilde{+}} \lk[
-\bar{\theta} \rho^0 \del_\tau \theta
+\bar{\theta} \rho^1 \del_\sigma \theta
+\frac{5}{2} p^{\widetilde{+}} \bar{\theta} \rho^0  \rho^1 \sigma^{5789} \theta
\re]+\II^{bos}_{PP}\ ,
\ee
where we have additionally fixed the worldsheet conformal symmetry 
by $h^{00}=1=-h^{11}$ and $x^{\widetilde{+}}=p^{\widetilde{+}}\tau$.
Hence, the $R\rightarrow \infty$ limit truncates away the 
non-linear terms in $\II^{(4)}$. 

Let us look at the mechanism of this simplification a bit closer. In terms of 
the length scale of the geometry $R$, the various objects of interest scale as
\bb\label{diman}
e^m_{0,1,5,6}\sim\frac{1}{R}\ \ ,\ \ 
e^m_{2,3,4,7,8,9}\sim 1\ \ ,\ \ 
G_{abcde}\sim \frac{1}{R}\ \ ,\ \ 
R_{abcd}\sim \frac{1}{R^2}\ .
\ee
This means that
\bb
V^\pm_i\sim R\ .
\ee
This is the central statement of the scaling limit. It trivially implies that the
kinetic term of the fermions scales as $R$ (see equation~\pref{Imainkin}).
And looking at the quartic terms~\pref{I4}, we can read off it that it
scales as $R^0$. Hence, in the limit $R\rightarrow\infty$, after rescaling
$\sqrt{R} \theta\rightarrow \theta$,
the quartic interactions vanish as $1/R^2$ while the kinetic terms are finite. 

The important point is that all
this actually follows from dimensional analysis and knowing the general
form of the action as given in~\pref{Imainkin},~\pref{I2} and~\pref{I4}. The five-form $G_{abcde}$ and 
the Riemann tensor $R_{abcd}$ (with tangent space indices)
must scale as $1/R$ and 
$1/R^2$ respectively as this is the only length scale in a
maximally symmetric geometry\footnote{
The term proportional to $DG$ in~\pref{R2} is zero since the five-form field
strength is covariantly constant when expressed in tangent space. 
}.
Similarly, the scaling of $V^\pm_i$ follows because of the same reason, {\em and}
the fact that our choice for light-cone directions did not involve either the holographic direction
$\rho$ or any coordinate on the five-sphere (see
equation~\pref{coord1} for why this matters). But in 
the derivation of the action~\pref{action}~\cite{SAHRR1}, these are
anyways needed assumptions from the outset to validate the expansion in 
superspace. Hence,
the disappearance of the quartic terms is a direct consequence of only dimensional
analysis and the fact that
our geometry has a single length scale; that is
given also knowledge of the general form of the action~\pref{action}.
In this sense, identifying other interesting scaling limits in other geometries is promising
as it may not require looking at any of the details of~\pref{action}.

Finally, it is easy yet considerably more nontrivial 
to identify which pieces quadratic in the spinors of~\pref{I2} survive the $R\rightarrow\infty$ limit. 
A short inspection of~\pref{AdSI2}, and using~\pref{diman},
picks out the two terms with four gamma matrix indices and
proportional to $\del_j\psi$.

\section{Discussion}

Given the complexity of the coupling of the RR fields to closed string dynamics, 
it is useful to reflect on all possible strategies of tackling this important problem.

\begin{itemize}
\item One approach consists
of attempting to get rid of the non-linear fermionic terms by choosing an appropriate ansatz for
the closed string dynamics, for example by focusing on left or
right moving modes. Solving this sector of the
theory first, we may hope to include additional vibrational
effects in a perturbative expansion around the ansatz.

\item We found that taking a scaling limit in a known geometry can be an 
easy way to
construct a simplified action. In particular, we realize that arguments such as
the one for the PP-wave geometry may follow from dimensional analysis and the general
form of the action~\pref{action}. This is hence an interesting direction to pursue. 

\item Another approach would look for special backgrounds for which the quartic terms vanish 
identically (at long wavelengths with respect to $\alpha'$). 
This route involves solving directly for
background field configurations such that the coefficients of the quartic terms in~\pref{cceq} are zero.
To illustrate the complexity of this task, let us look at
equations~\pref{pp} to~\pref{R2} set to zero, with the supergravity equations of
motion in mind. We may succeed in finding ansatz for which
these conditions 
involve only first derivatives of the fields.
We then use a flux configuration that satisifies the conditions
to determine the corresponding spacetime geometry. This approach appears rather involved and
contrived; but certainly a possible systematic strategy.

\item Another hope for computing with~\pref{action} would be to look for background configurations
that yield integrable non-linear sigma models. Indeed,
the fermionic part of our action, being at most quartic in the spinors, may be imagined to 
metamorphose, in certain special cases, into Gross-Neveu models~\cite{GROSSNEVEU}, with various 
symmetry structures for the quartic interactions. The integrability of the bosonic sector
% cite
is somewhat easier to establish (see for example~\cite{BCDLAX}). 
The guiding principle here is to
find a known candidate integrable system, and using its symmetries on the worldsheet, guess at the 
corresponding background geometry using the form of~\pref{cceq}.
We may then hope to find
examples of open/closed string duality such that both sides of the correspondence are
exactly solvable worldsheet theories, and the holographic duality amounts to an elaborate
`change of coordinate' or map from one to the other. This is a very interesting open problem that we 
hope to report on in an upcoming work. 

\item Lastly, one may try to identify a possible non-trivial vacuum for the spinors for
a given background geometry to leading order in a semiclassical approximation. This is
a rather complicated problem and it
appears the fruitfulness of this approach rests in the particular special features of the
background geometries one is to consider. 
\end{itemize}

In summary,
the most computationally tractable prospects for
learning from the action~\pref{action} involve restricting closed string dynamics to certain subsets of the
general dynamics; subsets or ansatz that circumvent the complicated problem of
understanding the effect of the quartic fermionic interactions. Short of this, one needs to 
hunt for exactly solvable systems or cascades of new scaling limits that
progressively simplify the problem.
It would also be interesting to study fluctuations of the closed string about the center of mass motion for
the NCOS case studied here. In particular, this would clarify the role of the interesting profile of the RR
axion flux in the geometry. 

\vspace{0.5in}
{\bf Note added:} When this work was in its final stages, a paper~\cite{MMS} appeared that overlaps
with part of the discussion in Section 4. 
\cite{MMS} rederives action~\pref{action} using a similar approach 
to~\cite{SAHRR1} without considering
the quartic spinor terms; this is mainly because \cite{MMS} focuses on PP-wave
backgrounds only, for which these terms vanish. In part of Section 4, our PP-wave action is obtained 
in a different approach. We point out that the process of scaling out the 
quartic terms in the PP-wave limit of the AdS geometry follows from dimensional analysis.

Finally, reference~\cite{MMS} states that
the derivation of~\cite{SAHRR1} involves gamma matrix manipulations, 
used also by~\cite{ATICKDHAR}, that they find
``confusing''. Indeed, an algebraic error can be identified in~\cite{SAHRR1} that implies that the action involves additional
quartic terms in the fermions that we have summed up in equation~\pref{cceq} as $\JJ$. Note that the action still 
truncates at quartic order, and has the general structural form indicated in~\cite{SAHRR1} and in equation~\pref{cceq}.
For the purposes of this work, this issue is not relevant since these additional terms have vanishing coefficients for the cases we study, as shown in the text. The interested reader is directed to~\cite{SAHRR1} where the full form of the action will be appropriately updated. 

\vspace{1cm}
{\bf Acknowledgements:} I thank Henry Tye for discussions. This work was supported in part by a grant from the NSF.

\section{Appendix A: IIB closed string action}

In this appendix, we collect the remaining pieces of the action given in~\pref{action} 
written in a slightly more convenient notation~\cite{SAHRR1}. This rewriting is straightforward
but algebraically involved, and we refrain from presenting any details.
In $\II^{(2)}$ given by~\pref{I2}, we have
\bbb\label{Aa}
\AA_a & \equiv  &
i \bar{\theta}\rho^1 \sigma_a^{\ b}\theta\ H^{(1)-+}_{\ b} 
-\frac{i}{4} \bar{\theta} \rho^1 \sigma^{bc}\theta\ H^{(1)}_{abc}
+\frac{i}{2} \bar{\theta}\rho^0\sigma^{bc}\theta\ \omega_{a,bc}
-\frac{i}{2} e^\phi \bar{\theta} \rho^0 \rho^1\theta\ \del_a \chi \nonumber \\
&-&\frac{i}{96} e^\phi\bar{\theta}\rho^0 \rho^1\sigma^{bcde}\theta\ G_{abcde}
+\frac{i}{12} e^\phi \bar{\theta} \rho^0 \rho^1 \sigma_a^{\ bcd}\theta\ G^{-+}_{\ \ bcd} \nonumber \\
&+&\frac{i}{4} e^\phi \bar{\theta} \sigma^{bc}\theta\ N_{abc}
-i e^\phi \bar{\theta} \sigma_a^{\ b}\theta\ N^{-+}_{\ \ \ b}
\eee
\bbb\label{Ba}
\BB_a & \equiv &
i \bar{\theta}\rho^0 \sigma_a^{\ b}\theta\ H^{(1)-+}_{\ b}
+\frac{i}{4} \bar{\theta} \rho^0 \sigma^{bc}\theta\ H^{(1)}_{abc}
-\frac{i}{2} \bar{\theta}\rho^1 \sigma^{bc}\theta\ \omega_{a,bc}
+\frac{i}{2} e^\phi \bar{\theta} \sigma_a^{\ b} \theta\ \del_b \chi \nonumber \\
&-&\frac{i}{2} e^\phi \bar{\theta} \sigma^{bc}\theta\ G^{-+}_{\ \ \ abc}
-\frac{i}{12} e^\phi \bar{\theta} \rho^0 \rho^1 \sigma_a^{\ bcd}\theta\ N_{bcd}
+i e^\phi \bar{\theta} \rho^0 \rho^1 \theta\ N^{-+}_{\ \ \ a}
\eee
$H^{(1)}_{abc}$ and $H^{(2)}_{abc}$ are respectively the NSNS and RR three-form field strengths.
In particular, we write as well $H^{(A)}=dB^{(A)}$; $e^\phi$ and $\chi$ are respectively the 
dilaton and RR axion fields; $G_{abcde}$ is the RR five-form field strength; and we have also defined
\bb
\NN_{abc}\equiv\chi H^{(1)}_{abc}+H^{(2)}_{abc}\ .
\ee
And $\omega_{a,bc}$ is the connection.
Note that all indices are in the tangent space and, in practice, need to be converted to spacetime
indices with the vielbein before making use of the action. The gamma matrices are
$16\times 16$  matrices\footnote{
Note that the metric signature used throughout is $(+--\cdots)$.}
\bb
\lk\{\sigma^a,\sigma^b\re\}=2 \eta^{ab}
\ee
acting in the space spanned by $\theta_A$.
And $\sigma^{a_1 a_2 a_3\cdots a_n}=\sigma^{[a_1} \sigma^{a_2} \sigma^{a_3} \cdots \sigma^{a_n]}$,
with antisymmetrization defined as in~\cite{SAHRR1}. 
Additional properties of the gamma matrices we use may be
found in~\cite{SAHRR1,HOWEWEST}. The light-cone gauge in spinor space is defined by
\bb
\sigma^+\theta_A=\frac{1}{2}\lk(\sigma^0+\sigma^d\re)\theta_A=0\mbox{     for  }A=1,2\ ,
\ee
$d$ being the chosen light-cone direction. Collecting the spinors in a doublet
$\theta\equiv \lk(\theta_1,\theta_2\re)$, we also define
\bb
\bar{\theta}\equiv \theta \sigma^- \rho^0\ ,
\ee
with $\rho$'s defined as
\bb
\rho^0=\lk(\begin{array}{cc}
0 & 1 \\
1 & 0
\end{array}
\re)\ \ \ ;\ \ \ 
\rho^1=\lk(\begin{array}{cc}
0 & 1 \\
-1 & 0
\end{array}
\re)\ ,
\ee
with the two-dimensional Dirac algebra
\bb
\lk\{ \rho^i,\rho^j\re\}=2 \eta^{ij}\ .
\ee
Our ten dimensional spinors $\theta_A$ acquire then a worldsheet spinor representation.

In $\II^{(4)}$, given by~\pref{I4}, we have
\bb\label{pp}
\PP\equiv
-\frac{1}{256} \lk(
e^{2\phi} G^{-+}_{\ \ \ abc} G^{-+abc}
+\frac{11}{2} e^\phi H^{-+T}_{\ \ \ a} \MM H^{-+a}
+\frac{29}{144} e^\phi H^T_{abc}\MM H^{abc}
+32 R^{-+-+}
\re)\ ;
\ee
\bbb\label{Q1}
\QQ^{(1)}_{ab} & \equiv &
\frac{1}{24} e^{2\phi} G^{-+}_{\ \ \ cda} G^{-+cd}_{\ \ \ \ \ \ b}
+\frac{1}{16} \lk(
\del_b \phi H^{(1)-+}_{\ \ \ \ \ \ a}
+e^{2\phi} \del_b \chi N^{-+}_{\ \ \ \ a}
\re) \nonumber \\
&-&\frac{29}{96} e^\phi H^{-+T}_{\ \ \ \ a} \MM H^{-+}_{\ \ \ \ b}
+\frac{1}{256} e^\phi H_{cda} \MM H^{cd}_{\ \ b}
+\frac{1}{2} R^{-\ \ +}_{\ \ a\ \ b}\ ;
\eee
\bbb\label{Q2}
\QQ^{(2)}_{ab} & \equiv &
\frac{1}{1536} \lk[
96 i e^\phi \del_b \phi N^{-+}_{\ \ \ \ a}
-96 i e^\phi \del_b \chi H^{(1)-+}_{\ a}
-392 i e^\phi H^{(1)-+}_{\ a} H^{(2)-+}_{\ b} \re. \nonumber \\
&-&\lk. 94 i e^\phi H^{(1)}_{cab} H^{(2)-+c}
+94 i e^\phi H^{(1)-+}_{\ \ \ \ \ \ c} H^{(2)c}_{\ \ \ \ ab}
+25 i e^\phi H^{(1)}_{cda} H^{(2)cd}_{\ \ b}
\re]\ ;
\eee
\bbb\label{R1}
\RR^{(1)}_{abcd} & \equiv &
\frac{1}{6144} \lk[
224 e^{2\phi} G^{-+}_{\ \ \ ead} G^{-+e}_{\ \ \ \ bc}
-208 e^{2 \phi} G^{-+}_{\ \ \ eab} G^{-+e}_{\ \ \ \ cd}
-96 \del_c \phi H^{(1)}_{abd} \re. \nonumber \\
&-&\lk.96 e^{2\phi} \del_c \chi N_{abd}
-336 e^\phi H^T_{bcd} \MM H^{-+}_{\ \ \ a}
+26 e^\phi H^T_{eac} \MM H^{e}_{\ bd}
+33 e^\phi H^T_{eab} \MM H^e_{\ cd} \re. \nonumber \\
&+&\lk. 384 R_{abcd}
\re]\ ;
\eee
and finally
\bbb\label{R2}
\RR^{(2)}_{abcd} & \equiv &
-i \frac{e^\phi}{3072} \lk[
384 D_a G^{-+}_{\ \ \ cdb}
+960 \del_a \phi G^{-+}_{\ \ \ cdb}
+192 \del_e \phi G^{-+e}_{\ \ \ \ cb} \eta_{ad}
+48 \del_c \phi N_{abd} \re. \nonumber \\
&-& \lk. 48 \del_c \chi H^{(1)}_{abd}
-4 H^{(1)-+}_{\ \ \ \ \ \ c} H^{(2)}_{abd}
+4 H^{(1)}_{abd} H^{(2)-+}_{\ \ \ \ \ \ c} \re. \nonumber \\
&-&\lk. 72 H^{(1)}_{ecd} H^{(2)e}_{\ \ \ \ ab}
+13 H^{(1)}_{eca} H^{(2)e}_{\ \ \ \ db}
\re]
\eee

We have defined the matrix~\cite{SL2Z}
\bb
\MM\equiv e^\phi 
\lk(\begin{array}{cc}
\chi^2+e^{-2\phi} & \chi \\
\chi & 1
\end{array}
\re)\ ,
\ee
and the doublet
\bb
H_{}\equiv \lk(
\begin{array}{c}
H^{(1)}_{} \\
H^{(2)}_{}
\end{array}
\re)\ .
\ee

While these equation are quite elaborate in form in the most general cases, for concrete examples
like the NCOS and AdS backgrounds we study in the text, they do collapse to much simpler forms.
And as a general rule, it is well-established that one is not to expect
life to be either simple or pleasant. 

The last term labeled $\JJ$ in equation~\pref{cceq} is also quartic in the fermions but involves combination that carry two
and four units of U(1) charge. These terms will be written down in explicit form in~\cite{SAHRR1}.

\section{Appendix B: NCOS background geometry}

We write in this appendix the NCOS background geometry 
used in Section 3. For more details, the
reader is referred to~\cite{VVSlargeM}. In the decoupling limit of interest, the string frame metric is given by
\bb
ds^2=\Omega^2 \lk(
\frac{v^2}{2\pi^2 \alpha_e}
\lk(dt^2-dy^2\re)
-\frac{dv^2}{v^2}-4 d\Omega_7^2
\re)\ ;
\ee
with
\bb
\Omega^2\equiv 8 \pi \alpha' \sqrt{\frac{G}{v}} \sqrt{1+G v^3}\ ,
\ee
where $G$ is the coupling of the dual NCOS theory, and $v$ is holographic coordinate 
identified with energy scale through the UV-IR relation. The 2-form gauge fields are given by
\bb
B^{(1)}_{ty}=\frac{\alpha'}{\alpha_e} G v^3\ ;
\ee
\bb
B^{(2)}_{ty}=-\frac{\alpha'}{\alpha_e} \frac{M}{(32\pi^2 G)^2}\ .
\ee
The dilaton is
\bb
e^\phi=\lk(32\pi^2\re)^2 \frac{G^{3/2}}{M} \frac{1+G v^3}{v^{3/2}}\ .
\ee
And the RR axion is given by
\bb
\chi=\frac{M}{\lk(32 \pi^2 G\re)^2} \frac{1}{1+G v^3}\ .
\ee
The $y$ coordinate is compact of size $\Sigma$. The NCOS theory is parametrized by
$G$, $M$, $\alpha_e$, and $\Sigma$. We take $M$ and $G$ large, and consider in particular
the scenario $M\rightarrow \infty$ with $G>1$ but finite. This yields to a finite theory where the
supergravity computations may be trusted. 

%\bibliography{biblio}
%\bibliographystyle{utphys}    

\providecommand{\href}[2]{#2}\begingroup\raggedright\endgroup

\end{document}